\documentstyle[10pt,aaspp4,flushrt]{article}
\def\etal{\it et al. \rm }
\begin{document}

\title{Gas Mass Fractions and the Evolution of LSB Dwarf Galaxies}

\author{James M. Schombert}
\affil{Department of Physics, University of Oregon, Eugene, OR 97403;
js@abyss.uoregon.edu}

\author{Stacy S. McGaugh} \affil{Department of Astronomy, University of
Maryland, College Park, MD 20742; ssm@astro.umd.edu}

\author{Jo Ann Eder} \affil{Arecibo Observatory\altaffilmark{1}, Puerto
Rico, 00612; eder@naic.edu}
\altaffiltext{1}{The Arecibo Observatory is part of the National Astronomy
and Ionosphere Center which is operated by Cornell University under
contract with the National Science Foundation.}

\begin{abstract}

The optical and H\,I properties for a sample of low surface brightness
(LSB) dwarf galaxies, cataloged from the Second Palomar Sky Survey, is
presented.  Gas mass fractions for LSB dwarfs reach the highest levels of
any know galaxy type ($f_g=$95\%) confirming that their low stellar
densities are due to inefficient conversion of gas mass into stellar mass.
Comparison with star formation models indicates that the blue optical
colors of LSB dwarfs is not due to low metallicity or recent star
formation and can only be explained by a dominant stellar population that
is less than 5 Gyrs in mean age.  If star formation occurs in OB
complexes, similar to normal galaxies, then LSB dwarfs must undergo weak
bursts traveling over the extent of the galaxy to maintain their LSB
nature, which contributes to their irregular morphological appearance.

\end{abstract}


\section{INTRODUCTION}

Dwarf galaxies are important to our understanding of galaxy formation and
evolution since they are the smallest sized, star-forming units at the
time of galaxy formation.  Under the bottom-up scenarios of galaxy
formation (Lake 1990, Baugh, Cole \& Frenk 1996), dwarf galaxies are the
building blocks to the entire Hubble sequence and, thus, the study of
dwarf galaxies is a look at the fossil remnants of the early Universe.  In
addition, dwarfs are extremely rich in dark matter compared to the amount
of known baryons calculated from starlight and neutral gas emission
(Ashman 1992, Carignan \& Purton 1998).  They serve as laboratories to
test dark matter candidates and the study of the formation and evolution
of the dark matter component in galaxies.  However, regardless of their
cosmological importance, the primary role for dwarf galaxies is that they
present us with the extreme limits with respect to star formation and
stellar populations in galaxies.  Much like the way an abnormal
psychologist studies extreme behavior to better understand normal
behavior, the study of dwarf galaxies leads to clues into the star
formation processes that are found in all galaxies, from dwarf to giant,
from quiescent to starburst.

The key to the star formation history of any galaxy, and its subsequent
evolution, is its gas supply.  Whether the gas supply is converted
completely into stars immediately after formation, as in ellipticals, or
whether the gas supply remains dispersed until a dynamic event increases
the number of cloud collisions, as in a tidally induced starburst, the gas
mass fraction, $f_g$, is the primary parameter for quantifying the
evolutionary state of a galaxy.  A galaxy's chemical and photometric
evolution closely follows the gas consumption rate.  For example, Bell \&
de Jong (2000) demonstrate that the metallicity of spirals follow a simple
closed-box chemical evolution model and also predicts the colors of the
underlying stellar population.  With respect to LSB galaxies, van den Hoek
\etal (2000) find that a majority of LSB disks can be explained by a
exponentially decreasing star formation rate ending with present-day gas
fractions near 0.5.

Historically, gas-rich galaxies have divided into two types, disk galaxies
and dwarf galaxies.  Disk galaxies are brighter and higher in surface
brightness and have dominated our studies of the star formation process.
In contrast, dwarf galaxies are fainter and lower in surface brightness,
making their detection and inclusion in galaxy catalogs problematic.
In the last decade, new galaxy catalogs (Schombert \& Bothun 1988, Impey
\etal 1996) have widened our range of central surface brightnesses to
include new extremes in low stellar densities (i.e. low surface
brightness, LSB).  The common interpretation is that these systems have
had, in the past, very low rates of star formation (de Blok \& van der
Hulst 1998) and, thus, there is the expectation that LSB galaxies should
be rich in gas compared to their stellar mass.  This was confirmed by
McGaugh \& de Blok (1997, hereafter MdB) in a study of a large sample of
disk galaxies over a range of surface brightnesses.  They found that there
is a strong correlation between a galaxy's gas supply and its stellar
density such that galaxies with the lowest surface brightness had the
largest $M_{HI}/L$ and $f_g$ ratios.  The gas fractions found by MdB also
indicated that LSB galaxies have the potential to become extremely bright,
high surface brightness (HSB) objects if some process increased the
efficiency of star formation and rapidly used the supply of gas (e.g. a
tidal interaction).  The discovery of star-forming dwarf populations in
the field (Driver, Windhorst \& Griffiths 1995, Schade \etal 1996) and in
distant clusters (Rakos, Odell \& Schombert 1997) makes an investigation
into nearby quiescent dwarfs timely.

The correlations found for disk galaxies have never been extended to dwarf
galaxies, primarily because of a lack of a uniform sample of gas-rich, LSB
dwarfs.  The purpose of this paper is undertake a similar analysis
performed by MdB on LSB disks to the LSB dwarfs from the PSS-II survey,
and to compare the results to the correlations found for all types of disk
galaxies.  The dwarf sample used herein is based on a visual search of
Second Palomar Sky Survey plates (Eder \etal 1989, Schombert, Pildis \&
Eder 1997), parallel to searches for large LSB galaxies (Schombert \&
Bothun 1988, Schombert \etal 1992).  This was primarily for a study on
biased galaxy formation, but optical and H\,I data is available for most
of the dwarfs.  The optical data for that sample was presented in Pildis,
Schombert \& Eder (1997). The H\,I data is presented in Eder \& Schombert
(2000), and forms the core of the analysis in this study.

\section{OBSERVATIONS}

The data for this paper is based on Second Palomar Sky Survey (PSS-II, see
Reid \etal 1991) plates cataloged in Schombert, Pildis \& Eder (1997).
The PSS-II differed from the original sky survey in that the latest Kodak
IIIa plates, which have greater resolution and depth than the original
surveys 103a emulsions (250 lines mm$^{-1}$ versus 80 lines mm$^{-1}$)
were used.  The plates used for the dwarf catalog are A or B grade,
selected for good surface brightness depth and covering declination zones
of the sky that can be observed with the 305m Arecibo radio telescope.

Dwarf candidates from the catalog were observed for the H\,I line at 21 cm
with the Arecibo 305m telescope during the 1992 and 1993 observing season.
All observations were made with the 21 cm dual-circular feed positioned to
provide a maximum gain (8 K Jy$^{-1}$) at 1400 MHz.  Total velocity
coverage of 8000 km s$^{-1}$ at a velocity resolution of 8.6 km s$^{-1}$
was used.  The observations were centered on 4000 km s$^{-1}$, which
avoided detection of the strong Galactic hydrogen signal on the
low-velocity end, and extended to 8120 km s$^{-1}$.  The HI observations
are reported in Eder \& Schombert (2000).  

Successful detections were later selected for follow-up CCD imaging on the
Hiltner 2.4m telescope located at Michigan-Dartmouth-M.I.T. (MDM)
Observatory.  We obtained images using either a Thomson 400 $\times$ 576
pixel CCD (0.25 arcsec pixel$^{-1}$) or a Ford-Loral 2048 $\times$ 2048
pixel CCD binned 3$\times$3 (0.51 arcsec pixel$^{- 1}$), with minimal
exposure times of 25 minutes in Johnson $V$ and 15 minutes in Johnson $I$.
The optical data  (luminosities, colors, scale lengths and surface
brightnesses is presented in Pildis, Schombert \& Eder 1997).  

In addition, data on ordinary spirals was extracted from two surveys,
Courteau (1996) and de Jong (1996), for comparison to the LSB dwarf data.
The Courteau survey was primarily focused on Sc galaxies, as probes to the
Tully-Fisher relation, presenting CCD and H\,I observations of 189
galaxies.  The CCD observations were obtained in $B$ and $R$, but a simple linear
transformation converts the $R$ magnitudes to $I$ to match our LSB dwarf
luminosities.  The de Jong sample consists of a range of spiral types,
imaged at $BVRIK$, selected from the UGC with diameters greater than two
arcmins and undisturbed in their morphology.  H\,I parameters were
obtained from a variety of sources in the literature using NED.

All distance related values in this paper use values of $H_o = 75$ km
sec$^{-1}$ Mpc$^{-1}$, $\Omega_o$ = 0.2, a Virgo central velocity of 977
km sec$^{-1}$ and a Virgo infall of 300 km sec$^{-1}$.  The data used for
this study is available at the LSB dwarf web site (zebu.uoregon.edu/$\sim$js).

\section{DISCUSSION}

\subsection{Optical Properties}

The optical properties of the PSS-II dwarf sample are presented in a
previous paper (Pildis, Schombert \& Eder 1997).  Unless specifically
mentioned, all luminosities are $I$ band values, scale lengths are
measured in $I$ band images.  The choice of $I$ band values is made to
minimize the error in calculating stellar mass from optical luminosities.
Studies by stellar population models (Worthey 1994) show that $M_*/L$
varies with time and star formation rate as a function of wavelength, but
is most stable in the far-red due to its distance in wavelength from the
region around the 4000\AA\ break.  Thus, $I$ band measurements 1) provide
a more accurate estimate of the stellar mass of a galaxy than bluer
passbands, 2) obtain a luminosity measure that varys less with recent star
formation than bluer passbands and 3) allows determination of structural
parameters (such as scale length) which are less distorted by recent star
formation events or dust.  Although the $I$ band is a difficult bandpass
to observe, due a bright background from atmospheric OH emission, grey and
bright time at many intermediate sized telescopes is much more accessible
than dark time, allowing long exposure times at bandpasses where the
moonlight has a minor contribution.

To place the LSB dwarf sample in context with respect to disk galaxies, we
have selected two comparison samples; the study of regular spirals by de
Jong (1996) and a study of Sc galaxies by Courteau (1996).  One advantage
to using the Courteau Sc disks and de Jong spirals is that both samples
were constructed using morphological criteria.  The Courteau Sc sample was
sorted from the NGC/UGC for late-type galaxies which display clear Sc-type
spiral structure and rotational symmetry, specifically to be homogeneous
for a study of the Tully-Fisher relation in late-type galaxies.  While
many of the Courteau objects are low in luminosity and small in size, they
are not dwarf-like in their appearance.  The de Jong spirals were selected
from the UGC with an emphasis on an undisturbed disk-like appearance and,
in this regard, compliments Courteau's Sc sample by covering a larger
range of disk Hubble types (from Sa to Sm).

The PSS-II dwarf galaxy sample was chosen on two primary characteristics;
dwarf-like (i.e. irregular) morphology and low surface brightness (see
Schombert, Pildis \& Eder 1997).  The morphology criteria was introduced
to maximize the cataloging of low mass objects to test theories of biased
galaxy formation, the original goal of the project, under the assumption
that all low mass objects have an irregular appearance.  This is not
always the case (e.g. dwarf ellipticals), but this criteria has the added
advantage of increasing the number of gas-rich dwarf galaxies in the
sample and the probability of detection at 21-cm.  The low surface
brightness nature of the sample is due to the fact that any search for
uncataloged galaxies will necessarily be limited by the depth of the
survey material.  Increased depth does not increase the detection of HSB
objects, which are already cataloged unless just below the angular size
threshold.  The greater sensitivity of the PSS-II plate material produces
an automatic bias towards LSB galaxies that were once too diffuse to be
visually detected and now have better contrast and, therefore, higher
visibility.

In a morphological sense, the de Jong and Courteau samples are
diametrically opposite to the PSS-II dwarf sample.  Where the disk samples
are selected for their regularity in spiral pattern, the LSB dwarfs are
selected with an emphasis on their irregularity.  The regularity of
morphology should be reflecting the star formation history of the galaxy,
where a regular pattern in the disk sample evolves from the ordered
motions of a density wave and, in contrast, the irregular nature of the
LSB dwarfs represents a chaotic history of star formation.  Additionally,
we can also compare the LSB dwarfs with dwarf sample of Patterson \& Thuan
(1996, hereafter PT); a sample of galaxies selected from the UGC by
morphology in a similar manner to the PSS-II sample and also imaged at
$I$.

\begin{figure}
\plotfiddle{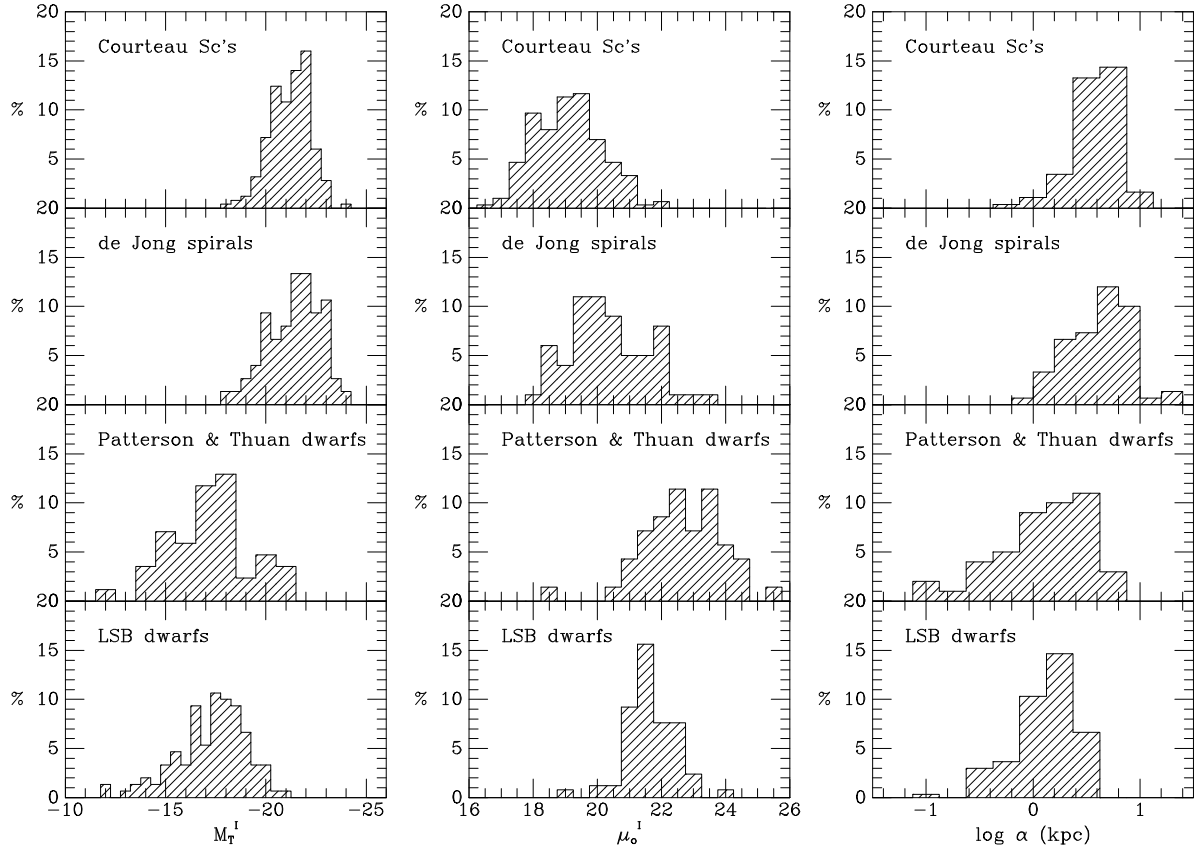}{9.0truecm}{-90}{65}{65}{-270}{360}
\caption{
Histograms of total luminosity ($M_T^I$), central surface
brightness ($\mu_o^I$) and scale length ($\alpha$) for the Courteau sample
of Sc galaxies, de Jong sample of ordinary spirals, Patterson and Thuan
UGC dwarf sample and the LSB dwarf sample.  The LSB dwarfs define a low
luminosity, low surface brightness and small scale length sample of
galaxies.
}
\end{figure}

A summary of the optical properties of the four samples is found in Figure
1; histograms of absolute magnitude ($M_T^I$), central surface brightness
($\mu_o^I$) and scale length ($\alpha$).  These values are taken directly
from Pildis, Schombert \& Eder (1997) and MdB.  The Courteau Sc sample was
converted from $R$ band data using his $B-R$ values and a linear
extrapolation to $R-I$ ($\langle{R-I}\rangle =0.5$ for late-type
galaxies).  Absolute luminosities are based on the total magnitude of the
galaxy (integrated curves of growth).  Central surface brightness and
scale length are based on exponential fits to the surface brightness
profiles.  Almost all the dwarf galaxies in this sample are well described
by an exponential surface brightness profile.  For objects with central
concentrations (bulge-like cores), the profile fit was made to the linear
region only, but the total magnitude includes the core luminosity.  Given
the extreme late-type nature of the samples, the core contribution is, in
any case, very minor.

There are several key points to note from the histograms.  One is that
there is significant overlap in all of the optical parameters between the
four samples.  In other words, there is no single property to a galaxy
where one could make a division and place dwarfs on one side and ordinary
disk galaxies on the other.  The de Jong and Courteau samples have similar
mean luminosities ($\langle{M_T^I}\rangle=-21.2$) which is much higher
than the LSB or PT dwarf sample ($\langle{M_T^I}\rangle=-17.4$).  This is
not surprising since spiral galaxies are known to have high rates of star
formation, often covering most of their disk area, resulting in more
stellar mass and, therefore, higher central surface brightnesses.  Surface
brightness is not directly correlated with total luminosity, as can be
seen from the distribution of the de Jong disk total magnitudes (see also
Driver \& Cross 2000).  The LSB dwarf sample displays a fairly sharp
cutoff in luminosity at $M_T^I=-19$, despite the fact that only morphology
and surface brightness were used to select the sample.

The trend of central surface brightness in each sample is as expected.
The LSB dwarfs have a low mean central surface brightness
($\langle{\mu_o}\rangle=21.8$), although the de Jong disk sample overlaps
the LSB dwarf sample (there are several Sd and Sm class spirals in the de
Jong sample).  The PT dwarf sample has a similar distribution with a
slightly fainter mean $\mu_o$.  The Courteau Sc sample has a brighter mean
$\mu_o$ ($\langle{\mu_o}\rangle=19.0$), but the range is quite broad.

The histograms of scale length, $\alpha$, are intriguing in that the de
Jong disk and Courteau Sc sample have identical distributions despite
having reached their optical appearances by very different star formation
histories (based on the final morphology and central surface brightness).
This leads us to conclude disk galaxies have the same range of sizes and
masses and Hubble class is imposed on disk galaxy structure by a set of
parameters only weakly linked to the structural ones (angular momentum,
environment, etc, see Zaritsky 1993).  The LSB and PT dwarfs have much
lower values of $\alpha$ with a cutoff at 3 kpc.  In fact, the size
distribution is perhaps the most accurate method of distinguishing dwarf
from ordinary galaxies when the mass is unknown (see Schombert \etal
1995).

\subsection{H\,I Properties}

It is often stated that LSB galaxies are gas-rich, which is loosely
defined to mean that they have high amounts of neutral hydrogen compared
to their optical luminosities.  It should be noted that to state that the
LSB dwarf sample is gas-rich does not imply that they have high H\,I
masses.  The distribution of H\,I masses (calculated from the total H\,I
flux using the prescription of Giovanelli \& Haynes 1988) is shown in
Figure 2 along with the H\,I mass distribution of the de Jong spiral,
Courteau Sc and PT dwarf samples.  The LSB dwarf distribution has a low
mean value ($\langle{M_{HI}}\rangle\, = 10^9 M_{\sun}$ compared to the
$\langle{M_{HI}}\rangle\, = 8\times10^9 M_{\sun}$ for the disk samples)
with a long tail towards low H\,I masses.  It is important to note that
dwarfs selected by optical characteristics, such as size, are also dwarfs
in terms of H\,I mass (see Eder \& Schombert 2000).

\begin{figure}
\plotfiddle{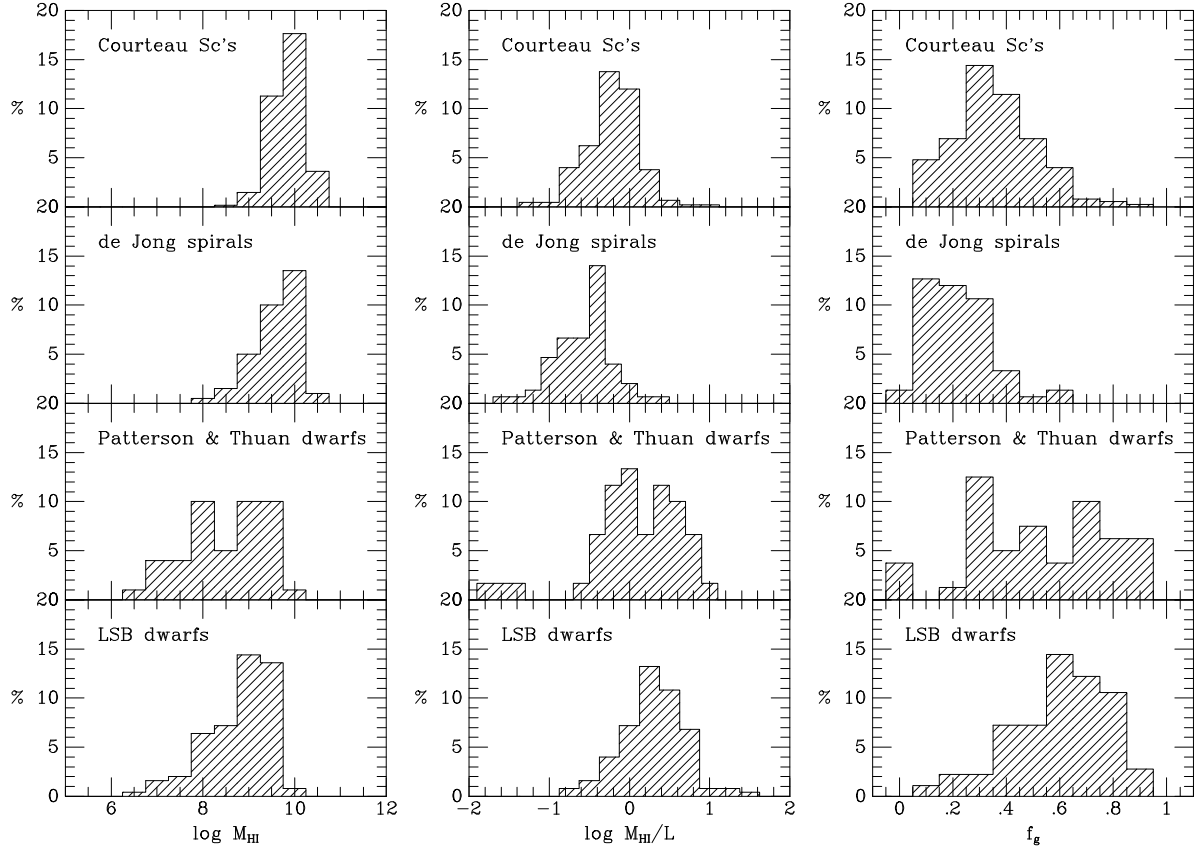}{9.0truecm}{-90}{65}{65}{-270}{360}
\caption{Histograms of H\,I mass, H\,I mass to light ratio ($M_{HI}/L$)
and gas mass fraction ($f_g$).  While LSB dwarfs are low in total gas
mass, their gas fractions are much higher than the spirals samples.}
\end{figure}

The dwarfs in Figure 2 with H\,I masses less than $10^8 M_{\sun}$ are of
interest to galaxy population studies since there is some suggestion of a
steeping of the H\,I mass function below $10^7 M_{\sun}$ (Schneider,
Spitzak \& Rosenberg 1998, Zwaan \etal 1997).  There is no correlation
between H\,I mass and surface brightness (Pildis, Schombert \& Eder 1997);
however, none of the $M_{HI} < 10^8 M_{\sun}$ dwarfs have central surface
brightnesses brighter than $\mu_o^I = 21$ (approximately $\mu_o^B =
22.5$).  This implies that low H\,I mass objects are under represented in
our catalogs, because of their bias towards high surface brightnesses
objects, and LSB, low H\,I mass galaxies have gone undetected in optical
surveys (Schneider \& Schombert 2000).

The distribution of the H\,I mass to luminosity ratio ($M_{HI}/L$) is
shown in the middle panel of Figure 2.  The disk samples cover the same
range (from $M_{HI}/L$ = 0.03 to 5); however, the Sc sample has a slightly
higher mean ($\langle{M_{HI}/L}\rangle$=0.8) than the de Jong spirals
reflecting the known dependence of $M_{HI}/L$ on Hubble type.  The LSB
dwarf sample has the highest mean of $M_{HI}/L$ values
($\langle{M_{HI}/L}\rangle$=3), reflecting the increasing importance of
neutral hydrogen to the baryonic content of these galaxies (see discussion
in \S3.4).  Matthews, van Driel \& Gallagher (1997) isolated a similar
sample of high $M_{HI}/L$ galaxies, also selected from late-type galaxies,
and the LSB dwarf sample share many of the same characteristics with their
galaxies.

\begin{figure}
\plotfiddle{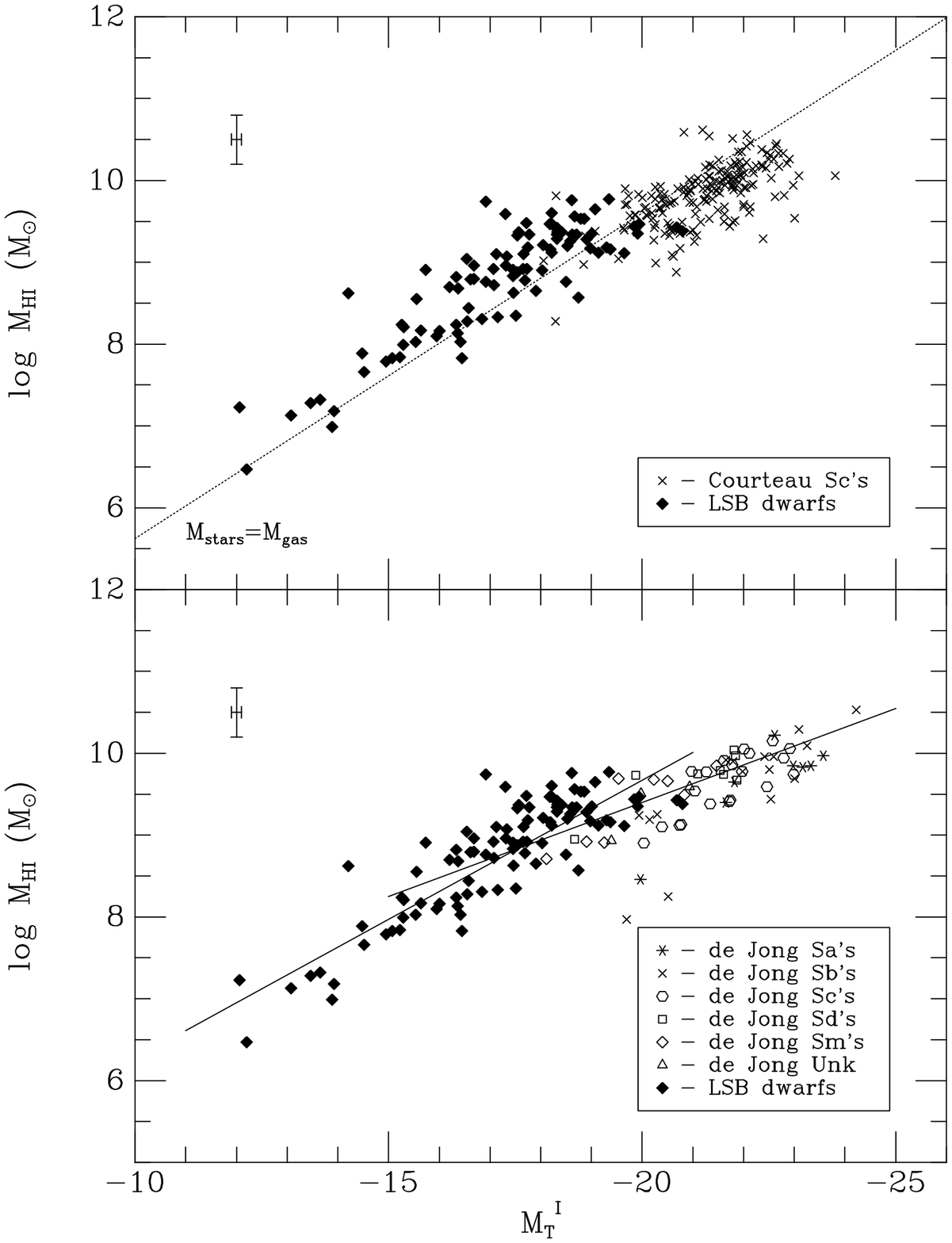}{14.5truecm}{0}{75}{75}{-240}{-50}
\caption{Total luminosity versus H\,I gas mass for LSB dwarfs compared
to Courteau Sc's (top panel) and de Jong spirals (bottom panel).  The line
of equal gas and stellar mass (assuming an $M_*/L$ of 1.7 (McGaugh \etal
2000) and a conversion of H\,I to gas mass of 1.4) is shown in the top
panel.  The ordinary spirals display a shallower slope in the gas to
stellar relationship, indicating that LSB dwarfs have been inefficient at
converting their gas into stars.}
\end{figure}

The relationship between stellar mass and gas mass is shown in Figure 3.
As discussed in \S3.1, the luminosity of a galaxy at 9000\AA\ is an
excellent measure of the number of stars in a galaxy since star formation
effects dominate in the blue portion of the spectrum.  To determine the
stellar mass of a galaxy, we require the mass to light ratio,
$\Upsilon_*$.  Ideally, we would like to measure $\Upsilon_*$ directly by
some dynamical means such as the observations of the vertical stellar
velocity dispersion (Bottema 1993).  Lacking such detailed information for
each galaxy, we follow the work outlined in MdB and de Jong (1996).  Based
on dynamical data and stellar population models (Bruzual \& Charlot 1993),
they determine that there is a factor of two spread in $\Upsilon_*$ in the
$B$ bandpass, but that in $I$ there is only a modest variation (less than
10\%).  Following the analysis presented in de Jong (1996), we adopt a
mean value of $\Upsilon_*$=1.7, although we note that a recent set of
spectroevolutionary models by Bell \& de Jong (2001) suggest that $M/L$
can vary as much as a factor of 2 over the color range of the dwarfs
presented herein (see their Figure 3).

The Courteau Sc sample is plotted with the LSB dwarfs in the top panel,
the de Jong spiral sample is plotted in the bottom panel.  Also shown in
the top panel is the equality line for gas and stellar mass, where the gas
mass is determined from the H\,I mass after corrections for metals and
non-atomic gas (see \S3.4).  There has been no detection of CO emission
from any LSB galaxy (Schombert \etal 1990, de Blok \& van der Hulst 1998),
so corrections for molecular gas are considered to be negligible (see
Mihos, Spaans \& McGaugh 1999).  For a majority of the disk galaxies in
the de Jong and Courteau samples, the baryonic matter is in the form of
stars since their data points lie below the $M_{stars} = M_{gas}$ unity
line.  The opposite is true of the dwarfs, where a significant fraction of
their baryonic matter is in the form of gas (primarily neutral hydrogen).
Thus, the term `gas-rich' dwarf refers to this reversal of stellar mass
dominance in disk galaxies to gas mass dominance for the dwarf sequence.

The correlation between H\,I mass and total luminosity is not linear (in
log space, i.e. different power law slopes) for the whole range of galaxy
luminosities.  In fact, the dwarf sequence clearly has a steeper slope
than both the de Jong spiral and the Courteau Sc sample.  Linear fits to
the three samples produce the following relations:

$$ {\rm log}\, M_{HI} = -0.22 M_T^I + 5.16\,\, {\rm (Courteau\,Sc's)}$$
$$ {\rm log}\, M_{HI} = -0.23 M_T^I + 4.80\,\, {\rm (de\,Jong\,disks)}$$
$$ {\rm log}\, M_{HI} = -0.34 M_T^I + 2.87\,\, {\rm (LSB\,dwarfs)}$$

\noindent where the fits to the disks and dwarfs are shown in the bottom
panel of Figure 3.  The error on the slopes is $\pm$0.02.  The de Jong
spirals are slightly more gas massive than the Sc sample at the low
luminosity end, in agreement with the findings of de Blok, McGaugh \& van
der Hulst (1996).  Some adjustment might be necessary to the Sc sample
since they contain a small fraction of their gas in molecular form
(although this amount is very small, see Young and Knezek 1989).

Assuming that $M_*/L$ does not systematically change with $M_{gas}$
between late-type disks and dwarfs, the relation between stellar and gas
mass is given by:

$$ M_{gas} \propto M_*^{0.55} {\rm (disks)}$$
$$ M_{gas} \propto M_*^{0.88} {\rm (dwarfs)}$$

\noindent The shallower slope for disk galaxies implies that they have
been more efficient at converting gas into stars in the past, assuming
that all galaxies form from a single reservoir of gas.  This also agrees
with the previous observation that disk galaxies typically have a greater
amount of their baryonic mass in stars rather than gas.  Interestingly,
even the LSB spirals in de Jong's sample, which have different star
formation histories from other spirals in the sample, display the same gas
to stellar mass behavior as the Sc galaxies.

\subsection{Gas Fractions}

The standard measure of the gas-richness of a galaxy is their ratio of the
gas mass to the luminosity, $M_{HI}/L$.  Figure 2 displays the
distribution of $M_{HI}/L$ for dwarfs versus the two spiral samples.  The
LSB dwarf sample has a higher mean $M_{HI}/L$ (5 versus 1/2 for disks),
but there is a great deal of overlap between the samples.  Note that many
LSB dwarfs presented herein have $M_{HI}/L$ values greater than 5.  This
represents new extremes in gas to light ratio since, for example, none of
the dwarfs studied by either van den Hoek \etal (2000) or van Zee, Haynes
\& Salzer (1997) have $M_{HI}/L$ values above 0.5 at $I$.  More relevant
is the trend of $M_{HI}/L$ with luminosity (i.e. stellar mass) and central
surface brightness (i.e. stellar density).  These diagrams are shown in
Figure 4 for the LSB dwarf and de Jong spiral sample.  Where the trend for
higher $M_{HI}/L$ with lower luminosity in the de Jong disk galaxies
continues to lower luminosities, the dwarf data, by itself, is not as
highly correlated as the disk galaxies.

\begin{figure}
\plotfiddle{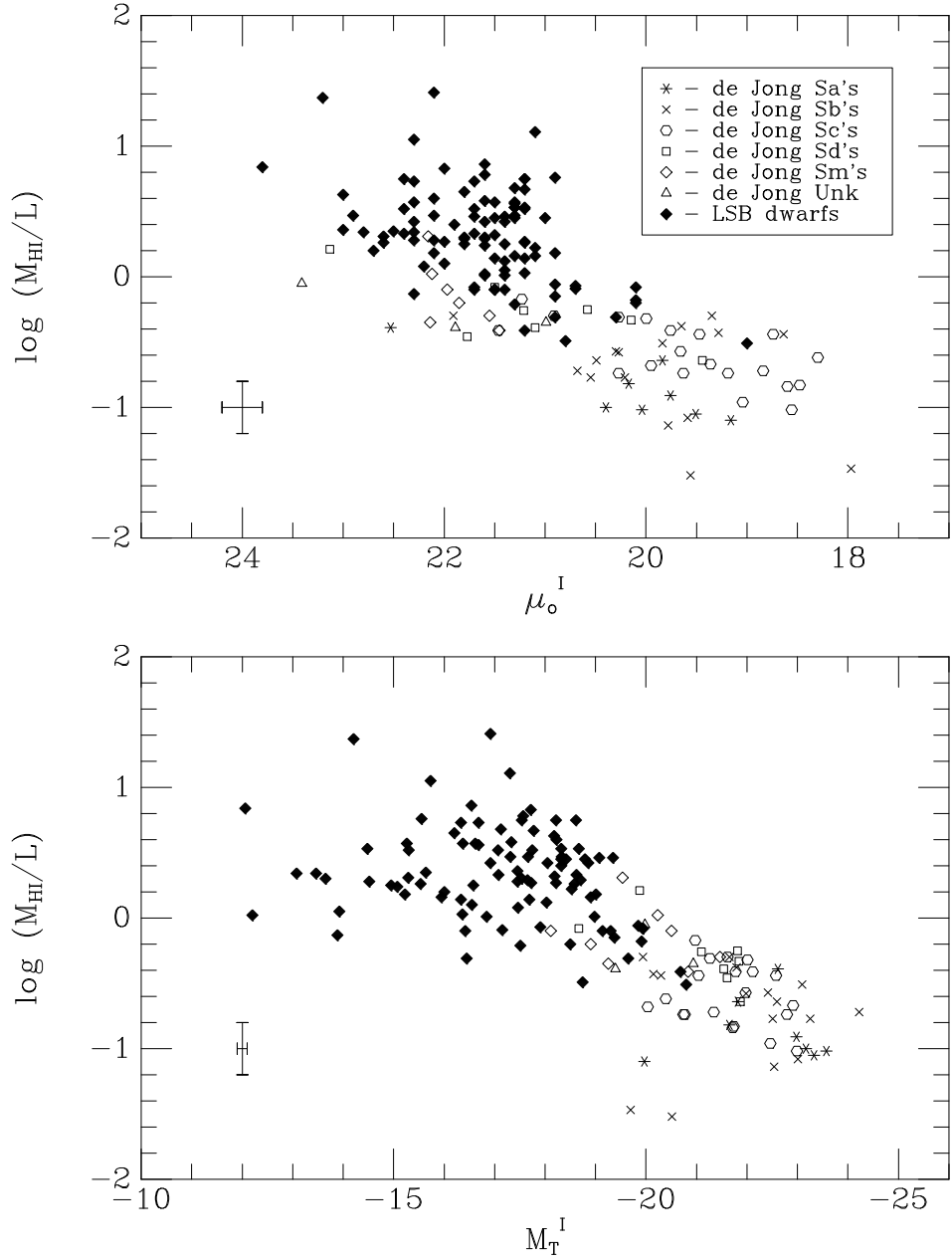}{14.5truecm}{0}{75}{75}{-240}{-40}
\caption{Gas to light ratios ($M_{HI}/L$) versus total luminosity and
surface brightness.  The previously strong correlations for disk galaxies
is extended by LSB dwarfs, but the LSB dwarfs are not as well-correlated
by themselves.}
\end{figure}

The change in the relationship between $M_{HI}/L$ and central surface
brightness ($\mu_o^I$) is particularly striking from the disks to dwarfs
(top panel of Figure 4).  The previously tight correlation for ordinary
spirals all but disappears for the dwarfs.  However, the dwarfs do serve to
fill the high $M_{HI}/L$, low $\mu_o^I$ region of the diagram.  In
fact, the previous correlations may be mostly due to various boundary
conditions imposed by our galaxy catalogs and unfilled regions of the
diagram which are empty for astrophysical reasons.  For example, the sharp
lower boundary may mark the limit placed by galactic winds from the first
epoch of star formation.  As the surface density of a galaxy drops (i.e.
lower surface brightness), it becomes easier to eject its ISM due to
heating by SN (Dekel \& Silk 1986).  Thus, the lower left region of the
$M_{HI}/L-\mu_o^I$ diagram would be inhabited by objects with neither
sufficient gas (due to blowout) nor stars (due to a lack of gas for star
formation) to be detected in the optical or radio (e.g., extremely LSB
dwarf ellipticals).  On the other hand, the upper right region of the
diagram is an area that would be occupied by extremely bright, high H\,I
mass galaxies.  Such objects are expected to be short-lived as it would be
unstable to strong star formation events, converting all its gas into
stars and, thereby, rapidly lowering its $M_{HI}/L$ value.

The $M_{HI}/L$ values can also be used to determine the gas fraction in a
galaxy.  As outlined by MdB, the baryonic gas mass fraction is given by

$$ f_g = {M_g \over M_g + M_*} $$

\noindent where $M_g$ is the total mass in the form of gas (neutral,
ionized and metals), and $M_*$ is the mass in stars.  To relate these to
the H\,I observations, we need the stellar mass to light ratio ($M_* =
\Upsilon_* L$) and the amount of gas represented by neutral hydrogen ($M_g
= \eta M_{HI}$). The parameter $\eta$ plays an analogous role for the gas
as $\Upsilon_*$ does for stars and accounts for both helium, metals and
for gas phases other than atomic hydrogen.  With these definitions, it is
straight forward to obtain

$$ f_g = \left( 1 + {\Upsilon_* L \over \eta M_{HI}} \right)^{-1}.$$

\noindent Again, following the analysis presented in McGaugh \etal (2000),
we adopt their mean value of $\Upsilon_*$=1.7.

For the gas, $\eta$ must be corrected for both the hydrogen mass
fraction $X$, and the phases of gas other than atomic.  We assume a
solar hydrogen mass fraction, giving $\eta = X_{\sun}^{-1} = 1.4$.
Variations in helium and metal content result in deviations from this
value of less than 10\%.  This is a very small effect compared to that
of other gas phases.  Ionized gas in H\,II regions and hotter plasma is
of negligible mass in late-type galaxies.  In addition, molecular gas
is also negligible for galaxy types later than Sc (Young \& Knezek
1989) and there has been no detectable CO emission in LSB galaxies
(Schombert \etal 1990).  Therefore, we adopt the solar hydrogen mass
fraction of 1.4 for $\eta$.

The histograms of gas fraction, $f_g$, for the LSB dwarf and comparison
disk samples are shown in Figure 2.  The gas fractions display similar
behavior to the H\,I mass distributions.  A majority of the galaxies in
the disk samples have $f_g$ values below 0.5; whereas, the dwarf galaxies
have very high $f_g$ values, many objects reaching an unprecedented 90\%
in gas fraction.  In the disk samples, the $f_g$ values peaked at 0.3, but
in the dwarf sample over 90\% of the galaxies have $f_g$ greater than
0.3.  Note that the high $f_g$ values for LSB dwarfs implies that there
has been no epoch of baryonic blowout such as has been found for
early-type galaxies (Bothun, Eriksen \& Schombert 1994, MacLow \& Ferrara
1999) unless there has been gas replenishment by infall, an unlikely
prospect given the low dynamical masses of these systems (Eder \&
Schombert 2000).  Adjusting the $\Upsilon_*$ for the bluer colors of LSB
dwarfs would increase, on average, the calculated $f_g$ making their
distribution even more extreme compared to disks.

The relationship between gas fraction and the optical properties of a
dwarf galaxy are not as strong as those for disk galaxies, but several
trends are clear.  Lower luminosity and low surface brightness (lower
stellar density) dwarf galaxies have much higher gas fractions than disk
galaxies.  Most dwarfs have higher gas fractions than either HSB or LSB
disk galaxies (see MdB Figure 8), yet lack the strong correlation with
luminosity or surface brightness that suggests a less orderly star
formation history compared to disks.  We interpret this trend to imply
that LSB dwarfs are LSB simply because they have not converted their gas
into stars at the same efficiency as either HSB or LSB disks.

To investigate this behavior further, we have adapted the chemical and
spectrophotometric model from Boissier \& Prantzos (2000, hereafter BP)
which present predictions of gas fractions, surface brightness and color
for a range of galaxy masses.  While intended to model disk galaxies, the
BP models follow a CDM framework and use scaling laws (Mo, Mao \& White
1998) that should be applicable to dwarf galaxies as well as disks.  The
smallest masses in their models correspond to rotation values of 80 km/s,
which is similar to the H\,I widths of the LSB dwarf sample.

A full description of the BP models can be found in their series of papers
(see BP for references).  Briefly, the models assemble a set of evolving
rings to represent a galaxy formed by primordial infall.  The CDM scenario
for galaxy formation requires the density fluctuations in the early
Universe give rise to dark matter dominated halos.  Within these halos,
baryonic gas condenses to form disks.  The resulting disks will have
characteristic central densities, scale lengths and masses in scalable
terms, although not necessarily correlated since a galaxy of a particular
mass can form with a range of central surface brightnesses and sizes.  In
order to distinguish various models, BP introduce the spin parameter,
$\lambda$, relating halo mass and angular momentum as described by Mo, Mao
\& White (1998), as the fundamental parameter to characterize the models.

Under this formalism, the spin parameter and circular velocity
(effectively, the disk mass) determine the characteristic timescale for
star formation.  In addition, the model galaxy's structure are such that
low $\lambda$ models ($\lambda$=0.01) simulate HSB galaxies and high
$\lambda$ models ($\lambda$=0.09) have surface brightness profiles that
recover that same structure as LSB dwarfs and disks.  The simulations also
allow the tracking of gas and stellar mass fractions as a function of
time.  By late epochs (13 Gyrs, see BP Figure 8), high $lambda$
simulations have gas fractions between 0.6 and 0.8, similar to the values
found for the LSB dwarfs in our sample.  All this formalism is normalized
to data for the Milky Way, thus completing the circle of structural and
stellar population effects.

\begin{figure}
\plotfiddle{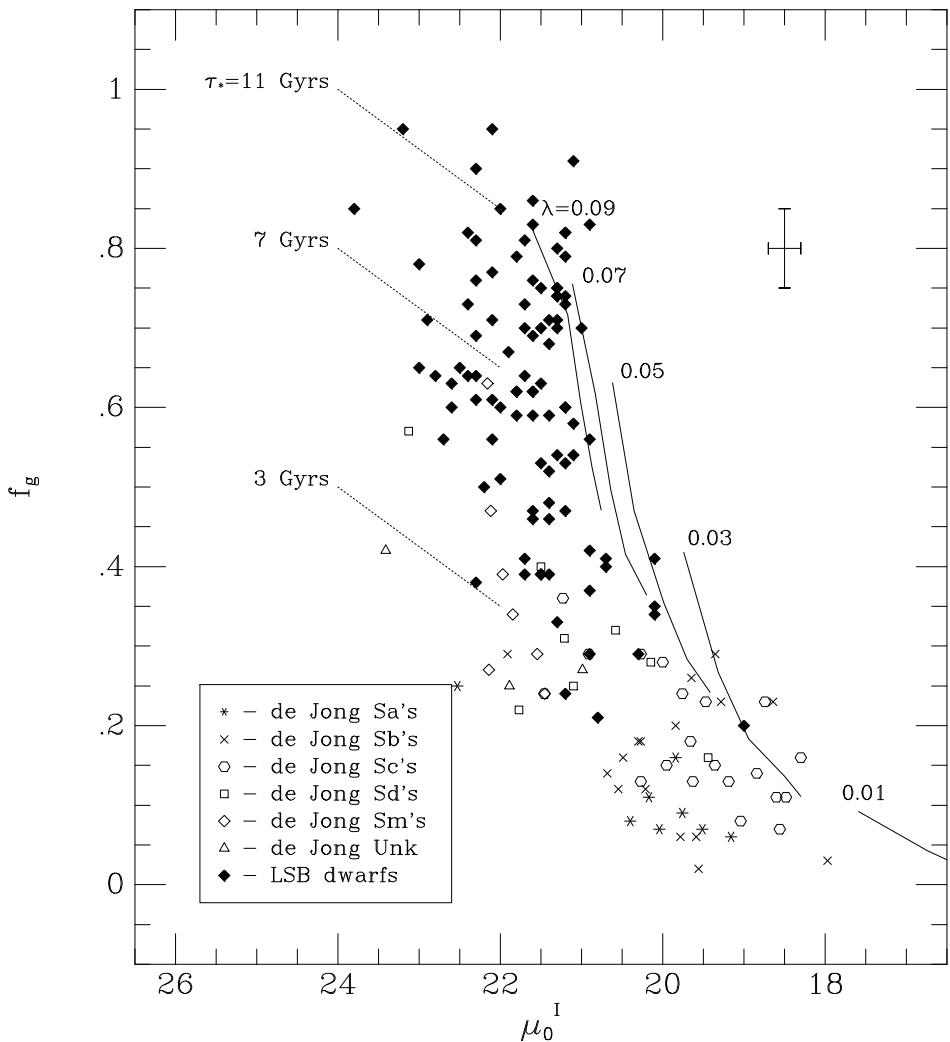}{14.5truecm}{0}{75}{75}{-240}{-40}
\caption{Central surface brightness ($\mu^I_o$) versus gas mass fraction
($f_g$).  The data from de Jong for disk systems is shown by morphological
type.  LSB dwarfs (solid diamonds) are, typically, much higher in gas
fraction.  Also shown are the 13 Gyr galaxy evolution models of Boissier
\& Prantzos (2000).  Each solid track follows the model gas fraction and
surface brightness as a function of mass for various spin parameters,
$\lambda$.  High $\lambda$ values correspond to LSB systems, low $\lambda$
values scale to normal disk galaxies.  Lower $f_g$ values on each track
indicate the higher mass models, opposite edge are for galaxies with
circular velocities of 80 km/s.  While the models predict the observed
$f_g$ values for LSB dwarfs, the model surface brightnesses are between
0.5 and 1 mags too bright.  Under the model formalism, this would imply
mean ages for the LSB dwarfs of 3 to 5 Gyrs younger than the shown
models.  Also shown are tracks of e-folding star formation timescales
(dotted lines) from the evolution models.  LSB dwarfs display longer star
formation timescales than disk systems.}
\end{figure}

A comparison of the BP models (at timestep 13 Gyrs) and the LSB dwarf gas
fractions and surface brightnesses is found in Figure 5.  Each set of
models for a specific spin parameter, $\lambda$, cover a range of gas
fractions and central surface brightness.  Each track represents a single
values of $\lambda$, where the brighter (lower $f_g$) edge of the tracks
represent high mass (high $V_c$) models and the fainter edge corresponds
to low mass systems.  As can be seen in Figure 5, the BP models predict
the general trend of higher $f_g$ with fainter $\mu_o$.  However, the
models appears to be about one magnitude too bright compared to the
position of the LSB dwarfs and disks.

A similar disagreement between models and observations was noted in BP for
the low surface brightness realm and it seems clear that additional models
with $\lambda > 0.1$ would begin to match the observed central surface
brightnesses of LSB dwarfs.  However, we also note that age could also
explain the discrepancy between the models and observations.  The low
mass, high $\lambda$ models typically brighten by one magnitude between 7
and 13 Gyrs (due to an increasing SFR and continued conversion of gas mass
into stellar mass).  Thus, if LSB dwarfs were, on average, 5 Gyrs younger
than other disk galaxies then the models would exactly match the data.
Color information confirms this interpretation as will be discussed in the
next section.  Since the lower $\lambda$ models require decreasing star
formation rates, then the surface brightness offset seen between the
models and the disk data is probably due to a mismatch in the models to
the accumulation of stellar mass relative to the Milky Way.  Younger age
is not an explanation for the disk systems since a decreasing star
formation leads to a fading in surface brightness, opposite to what is
seen in Figure 5.

The various $\lambda$ models also map into star formation e-folding times
(as given by Figure 5 in BP) as a function of galaxy mass.  Boundaries for
star formation timescales of 11, 7 and 3 Gyrs are also shown in Figure 5.
As one would expect from their high gas fractions, LSB dwarfs typically
lie in the region of the diagram occupied by galaxies with very long
e-folding timescales.  This would agree with their low metallicity values
(McGaugh 1994, van Zee, Haynes \& Salzer 1997) and the low current star
formation rates, based on H$\alpha$ imaging, and support the conclusion
that LSB dwarfs are slowly evolving systems.

\subsection{Colors}

The final clue to the star formation history of LSB dwarfs lies in their
colors.  Figure 6 displays the $V-I$ colors versus total luminosity
($M_T^I$), central surface brightness ($\mu_o^I$) and gas fraction ($f_g$)
for the LSB dwarfs and de Jong spirals.  $V-I$ is chosen as the comparison
color since it focuses on the mean color of the giant stars in a galaxy, a
measure of the star formation in the last 5 Gyr, rather than a color index
such as $B-V$, which is a measure of contribution from massive stars (i.e.
very recent star formation).  This is an important distinction since
interpretation using the $V-I$ colors can not discriminate between
constant star formation over 5 Gyrs or a series of short, weak bursts
within that same time frame.

\begin{figure}
\plotfiddle{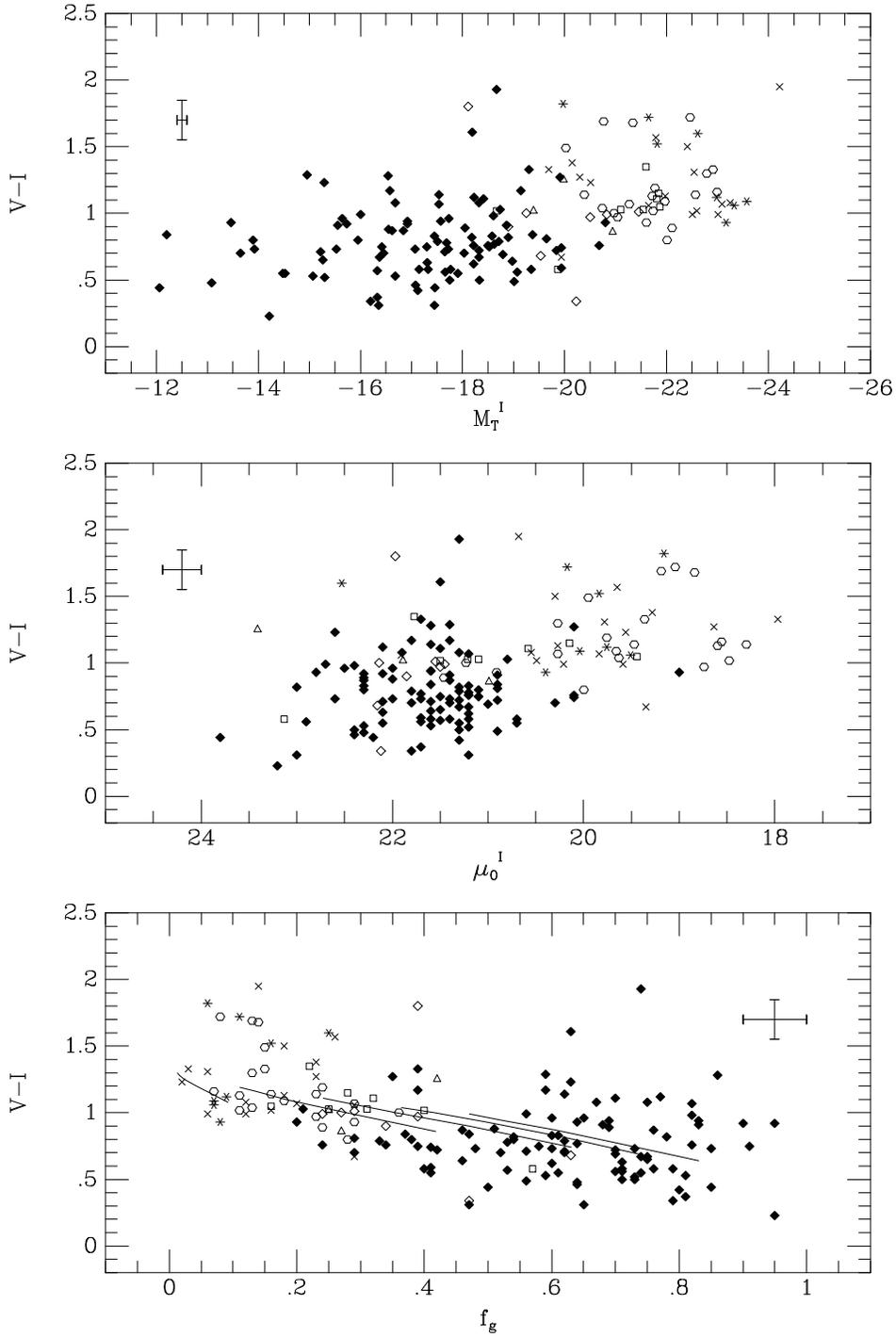}{17.0truecm}{0}{75}{75}{-240}{-20}
\caption{ $V-I$ color versus total luminosity ($M_T^I$), central surface
brightness ($\mu_o^I$) and gas fraction ($f_g$).  The LSB dwarfs continue
the sequence defined by ordinary spirals of fainter luminosities, higher
gas fractions and lower surface brightnesses with bluer colors.  The
models of Boissier \& Prantzos are shown in the bottom panel were high
$\lambda$ (LSB) are to the right. Symbols are the same as Figure 5.}
\end{figure}

Several familiar trends are evident from Figure 6.  One is that there is a
clear tendency for galaxies to have bluer colors with lower surface
brightness (middle panel, Figure 6).  This pattern was first noticed by
Schombert \etal (1992), and has been studied by several authors (Gerritsen
and de Blok 1999, Bell \etal 2000).  Blue optical colors were of early
importance since they eliminated the fading hypothesis for the evolution
of LSB galaxies (McGaugh \& Bothun 1994, see O'Neil \etal 1997 for the
discovery of red LSB galaxies).  It is interesting to note that the
scatter in the color diagrams is much less at $I$ versus the same diagrams
using $B$ or $V$ magnitudes (see McGaugh, Schombert \& Bothun 1997).  This
is due to the fact that LSB galaxies, by definition, have low luminosity
densities so that the contrast of recent star formation is extremely sharp
in blue indices, whereas the comparison of colors in the far-red tends to
average color changes due to short bursts of star formation.

McGaugh and de Blok found a strong correlation between color and magnitude
or surface brightness; however, despite the tendency for LSB galaxies to
have blue optical colors, there is no direct correlation between $V-I$ and
$M_T^I$ or $\mu_o$ for the LSB dwarf sample.  On the other hand, the
$V-I,f_g$ relationship for LSB dwarfs is a clear extension of the disk
relationships from MdB.  Interpretation of this trend is problematic.
Natively, one might expect that high gas mass fraction galaxies to have
red colors reflecting their low rates gas mass conversion into stellar
mass.  The Boissier \& Prantzos models are shown in the bottom panel of
Figure 6 and correctly predict the decrease in color with increasing gas
fraction.  This would seem to confirm several of the characteristics to
the high $\lambda$ models, such as a rising star formation rate and slow
chemical enrichment.  

Also visible in Figure 6 is that the dwarfs with the highest gas fractions
have the bluest colors.  Blue colors have always been an enigma in the
understanding of LSB stellar populations.  In general, the cause of the
bluer colors, compared to HSB galaxies, can either be 1) low mean
metallicity, 2) younger mean age or 3) recent burst of star formation and,
of course, some combination of the three.  For example, van den Hoek \etal
(2000) are able to reproduce the colors and gas fractions of LSB disks
using an exponentially decreasing star formation model which is consistent
with measured [O/H] values for their sample.  However, they are unable to
to fit the bluest, dwarf LSB galaxies in their sample without an
additional light contribution from a younger stellar population.

\begin{figure}
\plotfiddle{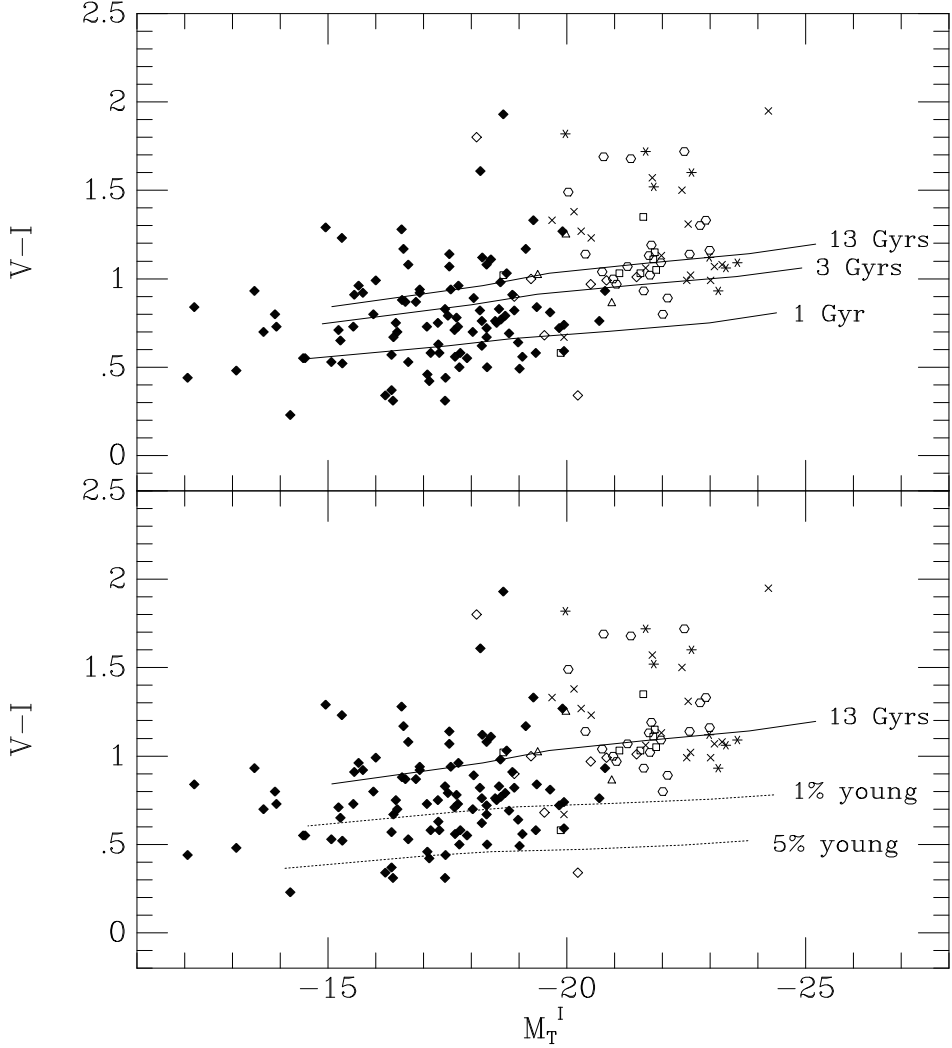}{13.0truecm}{0}{75}{75}{-240}{-40}
\caption{ $V-I$ color versus total luminosity ($M_T^I$). Symbols are the
same as Figure 6.  The top panel displays the multi-metallicity models of
Rakos \etal (2000) using the Bruzual \& Charlot (2000) SED's to convert to
$V-I$ colors.  Three population ages are shown; 13, 3 and 1 Gyrs.  A
majority of LSB dwarfs have mean $V-I$ colors which concur with mean ages
at least 10 Gyrs less than normal ellipticals.  The bottom panel displays
two `frosting' models where a fraction of a recent (100 Myr) population is
added to a 13 Gyrs population.  While only a 1\% burst is required to
match the colors of LSB dwarfs, this population would contribute 40\% of
the total light of the galaxy and eliminate its LSB appearance. Symbols are the same
as Figure 5.}
\end{figure}

Considering metallicity effects first, LSB galaxies are known to have
lower metallicities than HSB galaxies based on O[III] measurements of
H\,II regions (McGaugh 1994, van Zee, Haynes \& Salzer 1997).  These
values range from 1/3 to 1/20 solar, which corresponds to [Fe/H] of $-$0.4
and $-$1.3 dex.  While $V-I$ colors are not as sensitive to metallicity
changes as color indices such as $B-V$, a shift in 1.0 dex in [Fe/H] will
produce a change in 0.2 to the $V-I$ colors (Bruzual \& Charlot 2000).
Thus, some of the blueward slope in top panel of Figure 6 is due to a
decreasing mean metallicity from approximately solar at high masses to
1/25 solar at the faint end (Zaritsky 1993).

In order to examine the effects of metallicity, we have adapted the
multi-metallicity models for dwarf ellipticals from Rakos \etal (2000) to
the color-magnitude relation in Figure 6.  These models combine the
metallicity distribution given by the Kodama \& Arimoto (1997) infall
simulations with the single burst photometric models of Bruzual \& Charlot
(2000).  For each metallicity bin in the Kodama \& Arimoto model, the
fraction of stars are calculated and the flux, plus color, are determined
from the Bruzual and Charlot (2000) SSP's.  Then, the total integrated
color of a galaxy is calculated by summing the luminosity of each
metallicity bin population.  To simulate the change of metallicity with
mass, the shape of the Kodama metallicity distribution is held fixed but
the peak [Fe/H] is compressed to lower values but constant fractions.
This procedure successfully produces the correct slope and zeropoints to
the color-magnitude relation for ellipticals (see Rakos \etal 2000 for a
fuller discussion).

The resulting mass-metallicity relation for a 13 Gyr population, converted
into $I$ band magnitudes and $V-I$ color, is shown in the top panel of
Figure 7.  The model fits the blue edge of the de Jong disk galaxies
fairly well and is, of course, the mass-metallicity relation for spirals
documented by Zaritsky, Kennicutt \& Huchra (1994).  The number of disks
galaxies above the 13 Gyr line probably represents systems with reddening
due to dust.  We also note that few LSB dwarfs display an indication of
reddening, where the reddest LSB dwarfs are similar in $V-I$ color to
dwarf ellipticals.  This observation is in agreement with their lack of
IRAS detection (Schombert \etal 1990) and van den Hoek \etal (2000) also
rule out any significant extinction by dust in high gas fraction LSB
galaxies.

While the color-magnitude relation works well for LSB disks, HSB spirals
and ellipticals, a majority of the LSB dwarf sample continue to have $V-I$
colors too blue for the expectation from their metallicities.  Of course,
it is possible that the stellar population, that produces the optical
luminosity, has a much lower mean metallicity than the values measured
from nebular spectroscopy since those values represent the current
metallicity of ongoing star formation.  However, the multi-metallicity
models take into account the contribution from metal-poor stars and to
achieve the $V-I$ colors observed for LSB dwarfs would require a contrived
evolutionary history such that a majority of the stars have globular
cluster metallicities with a recent, and abrupt, rise in the mean
metallicity to match H\,II region spectroscopy.  Dominance of a globular
cluster metallicity population certainly can not be the case in all LSB
dwarfs since the galaxies with the highest $V-I$ values (the `red' edge in
Figure 7) follow the same mass-metallicity relation as disk and dwarf
ellipticals (predicting mean metallicities of [Fe/H]=$-$0.8 from the
models).  If the blueward dwarfs are due to metallicity effects, then the
mass-metallicity sequence breaks down below $M_T^I=-20$, which is not
indicated by LSB nebular spectroscopy (Bell \etal 2000).

An alternative explanation is that the spread in $V-I$ color is due to
changes in the mean age of the stellar population in LSB dwarfs.  To test
the effects of age, we have recomputed the luminosities and colors of the
multi-metallicity models for ages of 1 and 3 Gyrs (shown in Figure 7)
where an initial burst of star formation is assumed with a duration that
is less than 0.1 Gyrs.  The $V-I$ colors evolve quickly reaching the red
edge in only 5 to 6 Gyrs.  If mean age is responsible for the bluer $V-I$
colors, then an age between 1 and 3 Gyrs is indicated.  This, of course,
does not imply that LSB dwarfs formed less than 4 Gyrs ago since a young
mean age can be achieved through a number of scenarios.  For example, the
Boissier \& Prantzos models predict an increasing star formation rate with
time in their high $\lambda$, low mass simulations.  With e-folding
timescales between 7 and 11 Gyrs (as indicated by Figure 7), then a
majority of the luminosity in LSB dwarfs originates from stars with ages
less than 3 Gyrs even though there will exist stars dating back to the
epoch of galaxy formation.

A younger mean age can also be produced by a burst of recent star
formation on top of an underlying old population.  Recent star formation
can be tested for using so-called frosting models, where some percentage
of young stars is added to a 13 Gyr population.  The bottom panel of
Figure 7 displays the effects of adding 1\% and 5\% of a 100 Myr
population to the 13 Gyrs mass-metallicity sequence.  While a 5\% young
population is sufficient to match the $V-I$ colors of the LSB dwarfs in
the color-magnitude diagram, a star formation burst of this magnitude is
inconsistent with other properties of LSB galaxies.  For example, the
burst population in a 1\% population contributes 40\% of the total light
of the galaxy and in a 5\% population this percentage rises to 80\% of the
total light.  Given the slow rotation velocities for dwarfs, this young,
bright population would maintain a tight spatial correlation producing
many high surface brightness knots, in contradiction with the visual
appearance of the LSB dwarfs.  In order to maintain a low stellar density
and the lack of color gradient (Pildis, Schombert \& Eder 1997), the young
population would have to be evenly distributed instead of in cloud
complexes, which is style of star formation currently undocumented in
extragalactic studies.  In addition, even if the recent burst were spread
evenly within the LSB dwarf, the burst strength is inconsistent with the
current metallicities and gas fractions.

Lastly, the difference in $V-I$ colors between LSB dwarfs and disks may be
due to the fact that disk systems have high amounts of dust extinction.
Clearly some of the disk colors in Figure 6 are reddened, particularly the
objects between $V-I$=1.5 and 2.  Half of the early-type disks have $V-I$
colors greater than 1.2, the red edge for the LSB dwarfs, while the rest
of the same appear to be less effected by extinction effects and with a
blue edge to the disk sample near $V-I$=1.  Tully \etal (1998) finds that,
in the extreme case of a sample of edge-on galaxies, extinction at $I$ can
reach 1 mag, which would correspond to 0.3 in $V-I$.  However, studies of
overlapping pairs of ellipticals and late-type galaxies by Domingue, Keel
\& White (2000) find only modest amounts of extinction.  The correlation
of color-magnitude for disk galaxies near $V-I$=1 would argue that we are
observing a majority of the optical luminosity in disk galaxies unless
extinction effects conspire to redden the mean colors into this linear
relationship.

\section{CONCLUSIONS}

In this study we have isolated the optical and H\,I properties of a sample
of LSB dwarfs galaxies in order to study the relationship between gas
fraction and star formation history.  Our primary results can be
summarized as the following:

\begin{itemize}

\item[(1)] Objects selected by irregular dwarf-like morphology (to be
distinguished from irregular tidal morphology) do indeed define a sample
of galaxies that are faint, small and low mass (Pildis, Schombert \& Eder
1997).  However, no single characteristic defines a galaxy as uniquely
dwarf or giant.  There is a continuum of optical properties, such as
central surface brightness, luminosity and scale length, over the range of
galaxy Hubble types (Figure 1).

\item[(2)] The distribution of H\,I mass is such that the typical LSB
dwarf has less than $5 \times 10^9 M_{\sun}$ of neutral gas (Figure 2).
Over 1/3 of the dwarfs in this sample have H\,I mass less than $5 \times
10^8 M_{\sun}$.  A galaxy defined by irregular morphology will not only be
dwarf in size and luminosity, but also by H\,I mass and dynamical mass
(Eder \& Schombert 2000).

\item[(3)] There is a relationship between stellar mass and gas mass for
both dwarf and disk galaxies (Figure 3).  However, the relation for dwarfs
is much shallower then the relation for disks implying that disks have
been more efficient at converting gas into stars in the past (Bell \& de
Jong 2000).  This is in line with the more ordered appearance for disks
versus the chaotic structure for dwarfs, but this conclusion is dependent
on only small variations in $M/L$ between late-type disks and dwarfs,
which may not be the case (see Bell \& de Jong 2001).

\item[(4)] LSB dwarfs typically have much higher gas fractions (the ratio
of stellar to gas mass) than disk galaxies (Figure 2).  Many of the simple
correlations in disk galaxies between gas fraction and luminosity or
surface brightness are weaker in the dwarf realm.  LSB dwarfs with the
bluest colors have the highest gas fractions (Figure 6).

\end{itemize}

The gas mass fraction of a galaxy must be a key parameter to its
evolutionary path since the pattern, duration and strength of star
formation are directly determined by the amount of fuel available to
support star formation.  Disk galaxies, with red underlying stellar
populations and low gas fractions, suggests a star formation history that
involves a constant, or declining, star formation at relatively high rates
(Bell \& de Jong 2000, van den Hoek \etal 2000).  Under this scenario, the
gas fraction becomes a gauge of integrated past conversion of gas into
stars.  Thus, the Hubble sequence is a progression of star formation
rates, where Sa's had the highest rates and, in the present epoch, the
lowest gas fraction and Sc's have had lower rates and maintained higher
gas fractions.

For LSB dwarfs, their global characteristics are the reverse to ordinary
spirals.  They have very little in the way of a stellar population and a
majority of their baryonic mass is in the form of neutral hydrogen.  The
stellar population that does exist is very low in surface density and very
blue based on their $V-I$ colors (Pildis, Schombert \& Eder 1997).  Using
the prescription from van Zee, Haynes \& Salzer (1997), their integrated
past star formation rate is typically less than 0.1 $M_{\sun}$ yr$^{-1}$
and H$\alpha$ imaging of a few LSB dwarfs indicates current star formation
rates of 0.01 $M_{\sun}$ yr$^{-1}$.  On the other hand, their gas mass
fractions are very high, greater on average than the typical Sc galaxy.
Combining their current colors and star formation properties with the
knowledge that their metallicities are low (1/3 to 1/20 of solar), LSB
dwarfs must represent unevolved systems that have consumed very little of
their original gas supply in agreement with van Zee, Haynes \& Salzer
(1997) study of gas-rich dwarfs.

Our comparison to simple photometric/gas consumption models presented in
\S3.3 (displayed in Figure 5) provides the key difference in the
characteristics of dwarfs and disks.  Disk galaxies, over a range of
central surface brightnesses, have used a significant fraction of their
gas to produce stars (see also Bell \& de Jong 2000).  They have short
star formation e-folding times, but display relatively old mean stellar
ages reflecting a long history of substantial star formation.  Dwarfs, on
the other hand, have consumed very little of their gas due to their long
star formation timescales, and their current, dominant stellar population
has a very young age (less than 5 Gyr's on average).

Young mean age, either by constant or increasing star formation, appears
to be most likely solution to the optical colors presented in Figure 7.
With the levels of star formation deduced for the integrated past or for
the observed current, is difficult to envision star formation, spread
evenly in time and spatial position.  More likely, star formation in LSB
dwarfs proceeds in weak bursts that percolates over the spatial extent of
the galaxies with each event consuming only a small fraction of the local
HI gas mass.  This hypothesis of weak bursts on top of a very LSB stellar
population is in agreement with H$\alpha$ studies of LSB galaxies (Walter
\& Brinks 1999) where star formation traces both the gas and the surface
brightness and also in agreement with the conclusions of Bell \etal
(2000), who found a strong correlations with an LSB galaxies star
formation history and its local surface density.

We conclude with a comment on the differences between LSB dwarf galaxies
and LSB disk galaxies.  The star formation history of LSB disk galaxies
has recently been explored by both Bell \etal (2000) and van den Hoek
\etal (2000).  The study of LSB disks by van den Hoek \etal found that
they have gas fractions lower than the LSB dwarfs presented here, but
still higher than HSB disk galaxies (e.g., $f_g$ around 0.5).  They find
that exponentially decreasing SFR models are a good match to most of LSB
disk properties; however, weak bursts are indicated for the bluest
systems.  Bell \etal find similar results with the additional correlation
between local surface brightness (mass density) and past plus current star
formation rate.  Both studies find that LSB disks are lower in mean age
than HSB disks, mostly due to the more rapid build-up of a stellar
population in HSB with higher past rates of star formation.  In other
words, a higher surface brightness system has more old stars per pc$^{-2}$
and, thus, its calculated mean age from model comparison is older.  LSB
dwarfs have many characteristics in common with LSB disks, but the star
formation history can not be as smooth and uniform as proposed by Bell
\etal and van den Hoek \etal for LSB disks.  The disk correlations found by
MdB are much weaker for LSB dwarfs signaling a difference in the style of
star formation even within the class of LSB galaxies.  We suspect that
the dynamical state of disks versus dwarfs is responsible for the changes
seen in their stellar and gas properties as a galaxy transitions from
ordered rotation to the more turbulence dominated solid-body rotation found
in dwarfs (see van Zee, Haynes \& Salzer 1997).  However, that conclusion
remains in question until high resolution H\,I mapping of LSB dwarfs are
presented in a future paper.

\acknowledgements

We wish to thank the generous support of the Arecibo Observatory for the
allocation of time to search for HI emission from the candidate dwarf
galaxies and MDM Observatory in carrying out the photometry portion of
this program.  This work is based on photographic plates obtained at the
Palomar Observatory 48-inch Oschin Telescope for the Second Palomar
Observatory Sky Survey which was funded by the Eastman Kodak Company, the
National Geographic Society, the Samuel Oschin Foundation, the Alfred
Sloan Foundation, the National Science Foundation and the National
Aeronautics and Space Administration.

%

%
%
%
%
%
%

\end{document}